# Quantum Hall states observed in thin films of Dirac semimetal $Cd_3As_2$

Masaki Uchida,[1][*] Yusuke Nakazawa,[1] Shinichi Nishihaya,[1] Kazuto Akiba,[2]
Markus Kriener,[3] Yusuke Kozuka,[1] Atsushi Miyake,[2] Yasujiro Taguchi,[3]
Masashi Tokunaga,[2] Naoto Nagaosa,[1,3] Yoshinori Tokura,[1,3] Masashi Kawasaki[1,3]

[1]Department of Applied Physics and Quantum-Phase Electronics Center (QPEC),
the University of Tokyo, Tokyo 113-8656, Japan

[2]The Institute for Solid State Physics, the University of Tokyo, Kashiwa 277-8581, Japan

[3]RIKEN Center for Emergent Matter Science (CEMS), Wako 351-0198, Japan

[*]e-mail: uchida@ap.t.u-tokyo.ac.jp

**A well known semiconductor $Cd_3As_2$ has reentered the spotlight due to its unique electronic structure and quantum transport phenomena as a topological Dirac semimetal. For elucidating and controlling its topological quantum state, high-quality $Cd_3As_2$ thin films have been highly desired. Here we report the development of an elaborate growth technique of high-crystallinity and high-mobility $Cd_3As_2$ films with controlled thicknesses and the observation of quantum Hall effect dependent on the film thickness. With decreasing the film thickness to 10 nm, the quantum Hall states exhibit variations such as a change in the spin degeneracy reflecting the Dirac dispersion with a large Fermi velocity. Details of the electronic structure including subband splitting and gap opening are identified from the quantum transport depending on the confinement thickness, suggesting the presence of a two-dimensional topological insulating phase. The demonstration of quantum Hall states in our high-quality**

**Cd$_3$As$_2$ films paves a road to study quantum transport and device application in topological Dirac semimetal and its derivative phases.**

Topological materials, which are characterized by a non-trivial electronic band topology, have great potential for unprecedented quantum transport phenomena [1–5]. Among them, the topological Dirac semimetal (DSM) has attracted burgeoning attention as emergence of Dirac fermions in three-dimensional (3D) materials [1, 2, 6–14]. The 3D DSM state is particularly intriguing as a parent phase of exotic topological phases such as 3D topological insulator [3], Weyl semimetal [4], and two-dimensional (2D) topological insulator [5, 6], which are realized by symmetry breaking in DSM [1, 2, 7]. As crystalline materials of 3D DSM, Cd$_3$As$_2$ and Na$_3$Bi have been theoretically suggested [6, 8], and their key electronic structures have been directly confirmed by angle-resolved photoemission and scanning tunneling spectroscopy [7, 9–12]. A classification scheme of DSM in terms of the crystal point group symmetry has been also developed [13].

In this context, fabrication of DSM thin films is of crucial importance for exploring its potential as prototypical topological materials. Unlike other topological materials, however, it has been highly challenging to prepare high-quality DSM films. While Cd$_3$As$_2$ has been known as a stable II-V type semiconductor from the early period, its film quality has been limited due to the necessity of low-temperature growth for resolving its high volatility (Fig. 1a) [15–19]. Its electronic structure consists of conduction bands (CB) and valence bands (VB) with inverted orbital character, touching to form a 3D Dirac dispersion centered at the Dirac points $\pm k_\mathrm{D}$ (Fig. 1b). So far most of the transport studies including surface transport have been reported for bulk samples [20–25]. Tailoring confined Cd$_3$As$_2$ films with the Dirac dispersion thus opens up new avenues for research of quantized transport in this novel Dirac system such as by gate modulation [26].

Here we report the development of a growth technique to prepare high-quality Cd$_3$As$_2$ films and the observation of thickness-dependent quantum Hall effect. The films are epitaxially grown on an oxide substrate with accurately controlled thicknesses, yielding better crystallinity as compared to bulk single-crystals. As theoretically predicted for DSM [6], successive topological phase transitions to the 2D topological insulator and trivial insulator should occur through the quantum confinement (Figs. 1c and d), if the system becomes 2D in rather thick films. Owing to the large Fermi velocity of the Dirac dispersion, 2D quantum Hall states are actually observed up to such a thick ($\sim 23$ nm) film.

## Results

### Epitaxial film growth

High-quality $Cd_3As_2$ single-crystalline thin films are fabricated by combining pulsed laser deposition and solid phase epitaxy techniques (for details see Methods and Supplementary Figs. 2, 4, and 5). High-temperature annealing made possible by an optimized combination of double capping layers ($Si_3N_4$/$TiO_2$) and substrate ($SrTiO_3$) significantly improves the crystallinity and the electron mobility of the films. The film triangular lattice is epitaxially grown on the substrate square lattice with aligned in-plane axes, owing to a good match of their projected lattice distances. As confirmed in the transmission electron microscopy image (Figs. 1e and f), the Cd and As atoms are periodically arranged to form the low-temperature $Cd_3As_2$ structure without any discernible crystallographic defects. In the typical x-ray diffraction pattern (Fig. 1g), Bragg peaks of the (112)-oriented $Cd_3As_2$ film are observed with clear Laue oscillations consistent with the designed film thickness. The rocking curve of the film peak has a full width at half maximum of 0.02 degrees, which is sharper than typical values ($\sim 0.08$ degrees) reported for $Cd_3As_2$ single-crystals [20].

### Quantum transport measurement

Figure 2 summarizes high-field magnetotransport for a set of films with the same carrier density ($n = 1 \times 10^{18}$ cm$^{-3}$) and different thicknesses ($t = 12$, 14, 16, and 23 nm). Shubnikov-de Haas (SdH) oscillations and corresponding plateau-like structures are resolved from a few teslas in longitudinal resistance $R_{xx}$ and Hall resistance $R_{yx}$. As the field increases, integer quantum Hall states clearly emerge down to the quantum limit with filling factor $\nu = 2$. $R_{xx}$ is further suppressed and finally becomes zero. Simultaneously, $R_{yx}$ exhibits quantized values over wide field ranges, which are expressed as $1/R_{yx} = -\nu(e^2/h) = -sn(e^2/h)$, with the degeneracy factor $s$ and a non-negative integer $n$. A swell around the quantized values confirmed for thinner films is probably an artifact typically appearing in pulsed field measurements of such high resistance samples [27]. Although such deformations of transport data taken in pulsed fields were corrected by calculating the effective current through the sample as detailed in Methods and Supplementary Fig. 6, they cannot be completely removed so far. Absence of the half-integer plateaus suggests that a gap starts to open under the quantum confinement. Furthermore, the degeneracy factor $s$ shows a dramatic change depending on the film thickness, governing the appearance of the quantum Hall effect. It is altered from $s = 2$ to 1 when the thickness increases only by 2 nm from $t = 14$ to 16 nm. For clarifying the origin of this change, we analyse the quantum transport in detail.

The temperature dependence of the SdH oscillations was analysed for the whole series of Cd$_3$As$_2$ films, in order to extract effective masses and also quantum scattering times using the Dingle expression [28]. For the 12 nm film, as a typical example, the oscillation amplitude gradually decreases with elevating temperature but remains finite up to about 100 K (Fig. 3a). Its temperature dependence is suitably fitted to the standard Lifshitz-Kosevich formula (Fig. 3b), giving the effective mass of $m^* = 0.042m_0$. This light effective mass originating from the Dirac dispersion is in good agreement with values reported for bulk Cd$_3$As$_2$ [20–25].

A Landau-level fan diagram is plotted by following maxima and minima in the SdH oscillations (Fig. 3c). The slope dominant in the low-field region corresponds to the primary oscillation from the main Fermi surface as detailed later. In the thicker 16 and 23 nm films, on the other hand, the slope reduces almost by half above the critical field $B_c$, indicating a change in the degeneracy from $s = 2$ to $1$. This change is attributed to spin splitting, not to the lifting of other degeneracies, e.g., of valley or surface states. Quantum confinement is predicted to cause a change in the $g$ factor depending on the confinement thickness. In bulk $Cd_3As_2$, spin splitting of the oscillations is observed above $B \sim 10$ T [25], as in the thick 23 nm film, and the $g$ factor is estimated at $g \sim 15$ [12, 25, 29]. Reflecting the existence of other neighboring bands, the $g$ factor varies in inverse proportion to quadratic expression of the band gap $E_g$, according to the Roth equation derived in the second-order $k \cdot p$ perturbation theory [30]. The observed thickness dependence of the degeneracy can be thus understood from the rapid opening of the gap due to the confinement. Additionally, the Berry's phase $\phi_B$ can be estimated from the intercept in the fan diagram, based on the expression of the oscillating term [28]. The intercept $\gamma$ is typically about $-0.3$, which corresponds to a non-trivial Berry's phase of $\phi_B \sim 0.4\pi$, indicating the presence of relativistic Dirac fermions in the confined dispersion.

From the Fourier transformation of the SdH oscillations, further information about the 2D Fermi surface can be extracted (Fig. 3d). By applying the Onsager relation $A_F = (4\pi^2 e/h)B_F$ to the primary oscillation frequency $B_{F,1}$, the Fermi surface area $A_F$ is calculated to be $A_F = 3.3 \times 10^{-3}$ Å$^{-2}$ for the 12 nm film, for example. The dimensional change is also reflected in a clear field-angle dependence of the oscillation and magnetoresistance, as shown in Supplementary Figs. 8 and 10. The Fermi energy $E_F$, measured from the Dirac points, is estimated to be $E_F = 116$ meV by using $k_F = \sqrt{A_F/\pi} = 0.032$ Å$^{-1}$ and $v_F = \hbar k_F/m^* = 8.9 \times 10^5$ m/s in the following reported hyperbolic dispersion with an onset energy of $E_0 = 50$ meV and an energy difference between the conduction band bottom and the Dirac points of $E_{CB} = 35$ meV [6, 12].

$$E_F = \hbar v_F \sqrt{k_F^2 + (E_0/\hbar v_F)^2} - E_0 - E_{CB} \tag{1}$$

For the thicker 16 and 23 nm films, another peak $B_{F,2}$ is detected at lower frequencies, which is ascribed to the subband splitting due to the quantum confinement. The subband electronic structure is also evaluated assuming the same Fermi velocity and onset energy.

**Discussion**

The various quantum Hall states appearing in the $Cd_3As_2$ films can be comprehensively explained by considering a confinement effect on the original Dirac dispersion as schematised in Figs. 4a and b. For a more quantitative understanding, electronic band structures along the in-plane momentum direction ($k_{\perp[112]}$) and the film normal direction ($k_{\parallel[112]}$) are summarized in Figs. 4c and d. In Fig. 4c, $k_F$ and $E_F$ determined from the above analysis of the SdH oscillations using the dispersion relationship are plotted along the in-plane momentum direction. Here the band edge positions are interpolated from previous calculations [17]. The band gap $E_g$ is also almost consistent within the error bars with estimations from the $g$ factor change (for details see Methods).

The quantum confinement condition along the film normal direction is given by the following formula [31],

$$2k_{\parallel[112]}(E)t + \delta(E) = 2\pi n_c \tag{2}$$

or

$$k_{\parallel[112]}(E) = (2\pi n_c - \delta(E))/2t. \tag{3}$$

Here $t$ is the film thickness, $n_c$ is an integer numbering the confined subband, and $\delta(E)$ is the total phase shift at the interfaces (for details see Methods). In the structural plot expressing this relationship in Fig. 4d, the crossing point of the original dispersions ($k_{\parallel[112]}(E)$) and the quantization condition curves ($(2\pi n_c - \delta(E))/2t$) determines energy and momentum of the bottom of the subbands for each thickness. This agrees rather well with the experimental trends including the appearance of the second subband ($n_c = 2$) above 16 nm. Reflecting the large Fermi velocity of the Dirac dispersion, the band gap sharply opens when the confinement thickness decreases below 23 nm, giving rise to the dramatic $g$ factor change observed in the quantum Hall effect. The band character inversion, which occurs when crossing projected $k_D$, is also confirmed between $t = 12$ and 23 nm, as denoted by the CB character change from blue (Cd $5s$) to pink (As $4p$). In the case of the thick films where the subband is located inside the projected $k_D$, the gap energy and $g$ factor become nearly unchanged.

Novel topological phases derived from the 3D DSM state can be expected for the high-quality $Cd_3As_2$ films. As illustrated in Figs. 1b-d, for example, topological phase transitions to 2D topological (quantum spin Hall) insulator and trivial insulator, as proposed in the original theoretical work of topological states in $Cd_3As_2$ [6], should be induced by the confinement as long as the system remains 2D. Surprisingly, the two dimensionality is maintained up to 23 nm, where the orbital character only of the first subband ($n_c = 1$) is inverted, suggesting the presence of a 2D topological insulating phase at this thickness. Below this thickness, another topological phase transition to a trivial insulating phase occurs associated with the sharp change in the $g$ factor as confirmed in the thickness-dependent quantum Hall states. Since a magnetic field destroys the 2D topological insulating state by breaking time-reversal symmetry, nonlocal transport and scanning probe microscopy measurements are highly desirable for its further investigation. Applying electric gating, heterostructure fabrication, and chemical doping to such high-quality $Cd_3As_2$ films will open possibilities for further studying quantum transport and device application by tuning Fermi level, hybridization gap, and magnetic interaction in this system.

**Methods**

**Epitaxial film growth.**

While $Cd_3As_2$ has been known as a high-mobility semiconductor over half a century [32], its high-quality thin film growth has been quite challenging. In the 1970s and 80s, rather thicker films than 1 $\mu$m were grown by evaporating bulk $Cd_3As_2$ with a heater [15, 33–36] or a high-repetition-rate laser [16, 37], as plotted in Supplementary Fig. 1. More recently, since the topological Dirac semimetal state has been proposed for this system [6], a more elaborate approach using molecular beam epitaxy (MBE) has realized the epitaxial growth of single-crystalline thin films [17–19, 38]. Compared to bulk $Cd_3As_2$, however, their crystallinity and mobility are still limited due to the low-temperature growth. In this situation, we have developed and improved a high-temperature annealing technique with combining pulsed laser deposition (PLD), and obtained high-crystallinity and high-mobility epitaxial thin films comparable to bulk quality. The mobility reaches maximum of $\mu = 3 \times 10^4$ cm$^2$/Vs for $n = 1 \times 10^{18}$ cm$^{-3}$, while it intrinsically decreases with reducing dimensions (thickness) from three ($\gtrsim 80$ nm) to two ($\lesssim 40$ nm).

$Cd_3As_2$ films and $TiO_2$ / $Si_3N_4$ capping layers were deposited using KrF excimer laser. Optimization of the capping materials and their combinations has enabled high-temperature annealing of the film, while the idea of adopting a protective layer was already tried for $Cd_3As_2$ growth [15]. A $Cd_3As_2$ polycrystalline target was prepared by mixing 6N5 Cd and 7N5 As shots at a ratio of 3:2, keeping the mixture at 950 °C for 48 hours in a vacuum-sealed silica tube, grinding and pelletizing the compound, and then resintering it at 250 °C for 30 hours. The three layers were successively deposited on (001) $SrTiO_3$ single-crystalline substrates, at room temperature and below a base pressure of $10^{-7}$ Torr. Typical laser conditions (fluences, repetition rates) were (0.6 J/cm$^2$, 10 Hz), (4 J/cm$^2$, 20 Hz), and (4 J/cm$^2$, 20 Hz), for $Cd_3As_2$, $TiO_2$, and $Si_3N_4$, respectively. The shape of the Hall bar, as shown in the inset of Fig. 2d, was defined by employing a stencil metal mask for the successive deposition of the $Cd_3As_2$ and $TiO_2$ layers, while the $Si_3N_4$ layer was then deposited on the entire substrate to cover the Hall bar edges. After annealing the sample at 600 °C in air, an (112)-oriented $Cd_3As_2$ film is formed through epitaxial crystallization.

Supplementary Fig. 2 demonstrates the annealing effect probed by x-ray diffraction (XRD). A sample consisting of $Cd_3As_2$ (14 nm) and $TiO_2$ (30 nm) / $Si_3N_4$ (200 nm) layers shows only the $SrTiO_3$ substrate peaks before annealing. After annealing the sample at 600 °C in air, in stark contrast, (112)-oriented $Cd_3As_2$ film peaks become clearly evident through epitaxial crystallization. Three different periodic components from the respective layers remain observed in both the Kiessig fringes. Many combinations of other substrates ($Al_2O_3$, $BaF_2$, $CaF_2$, InP, CdTe, mica) and capping materials (Cr:$Al_2O_3$, $SiO_2$, MgO, $CaF_2$, Si) were also tested, but they resulted in chemical reaction to substrates, cracking of capping layers, or poor crystallinity of the films. As shown in Supplementary Fig. 3, $TiO_2$ / $Si_3N_4$ capping layer conduction, which is probably due to slightly oxygen-deficient $TiO_2$ deposited under the high vacuum, is more dominant at high temperatures above about 100 K, particularly for thinner $Cd_3As_2$ films with higher resistance. At low temperatures, on the other hand, the capping layers become highly insulating and only the $Cd_3As_2$ film conduction remains, ensuring intrinsic quantum transport measurements of the $Cd_3As_2$ films.

Detailed XRD data of the obtained epitaxial film, confirming its high crystallinity and flatness, are summarized in Supplementary Fig. 4. A magnification around the (224) film peak shows clear Laue oscillations consistent with the designed thickness. A rocking curve of the peak has a full width at half maximum (FWHM) of 0.02 degrees, which is sharper than the typical value (0.08 degrees) reported for bulk $Cd_3As_2$ single-crystals [20]. A $\phi$ scan with twelve-fold symmetry reveals that the in-plane $[1\bar{1}0]$ axis is exactly aligned with the [100] or [010] axes in the substrate, depending on two possible stacking patterns of the six-fold triangular lattice on the substrate square lattice. One reason of the successful epitaxial growth is probably that the projected lattice distance of the $Cd_3As_2$ layer (3.88 Å) has a good match with the $SrTiO_3$ one (3.91 Å). The domain size is estimated at about a few tens of microns from the STEM observations on various areas, which are comparable to the channel length scale ($\sim 60$ $\mu$m) but much larger than the magnetic length $l_B = \sqrt{\hbar/eB}$ ($\sim 26$ nm at 1 T).

Atomic-scale images of the $Cd_3As_2$ single-crystalline film are displayed in Supplementary Fig. 5, taken with cross-section high-angle annular dark-field scanning transmission electron microscopy (HAADF-STEM) and energy dispersive x-ray spectrometry (EDX). Cd and As atoms are periodically arranged without any clear crystallographic defects over a wide area. A shift of Cd atoms present in the low-temperature phase is also detected in the magnified image. From this view direction, it is difficult to determine whether the originally proposed ($I4_1cd$) [39] or the recently corrected ($I4_1/acd$) [40] structure is formed in the film, while both have the similar electronic structure.

**Quantum transport measurement.**

Transport measurements up to 55 T were performed using a nondestructive pulsed magnet with a pulse duration of 37 ms at the International MegaGauss Science Laboratory at the Institute for Solid State Physics of the University of Tokyo. Longitudinal resistance $R_{xx}$ and Hall resistance $R_{yx}$ were measured on the 60 $\mu$m-width multi-terminal Hall bar with flowing a DC current of $I = 5$ $\mu$A. In this Hall bar configuration, unexpected effects on the transport such as the current jetting effect in high-mobility semimetals are avoided. Aluminum electrode wires were connected to the Hall bar edges by using an ultrasonic bonding machine and then their connections were reinforced by applying silver paste. Small deformations of transport data taken in the pulsed magnetic fields were corrected based on a simple classic model [27]. In general, when measuring a resistive sample in pulsed fields, a small capacitive component connected parallel to the sample is slightly charged or discharged depending on the resistance change, leading to non-negligible time variation of an effective current through the sample. With the increase in the sample resistance $R_{xx}$, the effective current $i_x$ deviates more from the original set current $I$, and the deformations become more serious. By numerically solving the following differential equation detailed in ref. [27] with a capacitance of $C = 4$–9 nF, we could calculate the exact current $i_x$ and obtain data showing negligibly small hysteresis between forward and backward field sweeps, as exemplified in Supplementary Fig. 6.

$$\frac{di_x}{dt} = \frac{I - i_x(1 + CdR_{xx}/dt)}{R_{xx}C} \qquad (4)$$

Supplementary Fig. 7 demonstrates $R_{xx}$ for the 12 nm $Cd_3As_2$ film, measured from 1.4 to 50 K in the pulsed high fields. $R_{xx}$ minima at the $\nu = 2$ quantum Hall state slowly increase from zero with elevating temperature, which can be well fitted with the standard Arrhenius plot. Obtained high activation energy of $\Delta = 19$ K is ascribed to the unusually high Fermi velocity in $Cd_3As_2$.

At low fields, $R_{xx}$ and $R_{yx}$ were measured using a Quantum Design Physical Properties Measurement System cryostat equipped with a 9 or 14 T superconducting magnet. Supplementary Fig. 8 plots the data in the low-field region for films of various thicknesses, showing clear Shubnikov-de Haas (SdH) oscillations and Hall plateaus from a few teslas. The degeneracy factor $s$ in the quantization formula can be extracted from the increment of the plateau values. Change of the degeneracy from $s = 2$ to 1 is observed for the 23 nm film at low fields, indicating that spin splitting of oscillations occurs above about 12 T. Apparent degeneracy of $s = 4$ observed for the 16 nm (from $\nu = 16$ to 12) and 23 nm (from $\nu = 20$ to 16) films is ascribable to subband crossing. Corresponding beating pattern due to existence of the another subband can be also confirmed in $R_{xx}$ for the 16 nm and 23 nm films. In contrast to the films below 23 nm, the Hall plateaus in $R_{yx}$ become much less pronounced in the 37 nm film and almost completely disappear for the 100 nm film. This suggests that the system gradually changes from two-dimensional (2D) to three-dimensional (3D) around 40 nm.

Supplementary Fig. 9 compares temperature dependence of the SdH oscillations and their analysis to extract effective mass ($m^*$) and quantum scattering time ($\tau_q$) following the Dingle expression [28].

$$\frac{\Delta R_{xx}}{R_0} \propto \frac{4\zeta}{\sinh\zeta}e^{-\pi/\omega_c\tau_q}, \zeta = \frac{2\pi^2 k_B T}{\hbar\omega_c} \tag{5}$$

Here $\Delta R_{xx}/R_0$ is the oscillation amplitude normalized by the zero-field resistance and $\omega_c = eB/m^*$ is the cyclotron frequency. To investigate the main conduction band, the analysis is performed assuming a single band, although the oscillation amplitude at each Landau index is affected by the existence of other subbands for the 16 and 23 nm films. The effective mass is slightly decreased with decrease of the confinement thickness, probably due to the dispersion curvature change associated with the gap opening.

Supplementary Fig. 10 shows angular-dependent SdH oscillations in $R_{xx}$, measured also with a conventional superconducting magnet. When the applied magnetic field is tilted from out-of-plane ($\theta = 0°$) to in-plane ($90°$) direction in the 23 nm film, the oscillation period as well as the amplitude is substantially reduced. A gradual dimensional change from a cylindrical (2D) Fermi surface to a spherical (3D) Fermi surface is observed between 23 and 100 nm, consistent with the thickness dependence of Hall plateaus in Supplementary Fig. 8. For the 37 nm film, a cylindrical but corrugated (quasi-2D) Fermi surface is confirmed in the beginning of the dimensional change. Weak corrugation can be confirmed also for the 23 nm, but the 23 nm film rather closer to 3D shows much higher second frequency $B_{F,2}$ than the 16 nm one, eliminating the possibility of the neck orbit as a cause of $B_{F,2}$. Along with this dimensional change, considerably large negative magnetoresistance probably due to so called chiral anomaly [14] is also observed for the $B \parallel I$ configuration ($\theta = 90°$) on the Hall bar.

Landau-level fan diagrams magnified around the origins are shown in Supplementary Fig. 11 for all the thicknesses. The Berry's phase $\phi_B$ can be estimated from the intercept $\gamma - \delta$ in the fan diagram, on the basis of the following expression of the oscillating term in $\Delta R_{xx}$ [28].

$$\frac{\Delta R_{xx}}{R_0} \propto \cos[2\pi(B_F/B - \gamma + \delta)] \tag{6}$$

Here $B_{\mathrm{F}}$ is the SdH oscillation frequency, $\gamma$ is the phase factor expressed as $\gamma = 1/2 - \phi_{\mathrm{B}}/2\pi$, and $\delta$ is the phase shift being zero in two dimensions. For the 12 and 14 nm films, the intercept is about $-0.3$, which corresponds to the non-trivial Berry's phase of $\phi_{\mathrm{B}} \sim 0.4\pi$. In Cd$_3$As$_2$, the Berry's phase extracted from the intercept has been highly scattered and controversial [17, 20, 21, 24, 25, 41–43]. According to the recent theoretical calculation [43], when the Fermi energy is located above the saddle point of the two Dirac dispersions as in the cases of the previously reported carrier densities, the non-trivial phases at $\pm k_{\mathrm{D}}$ are cancelled out ($\phi_{\mathrm{B}} = \phi_{\mathrm{B},+k_{\mathrm{D}}} + \phi_{\mathrm{B},-k_{\mathrm{D}}} = 0$). In our two-dimensional case, however, the non-trivial Berry's phase remains finite without the cancellation within the confined subband, indicating the presence of the relativistic Dirac fermions therein. For the 16 and 23 nm films, modulated behavior due to formation of the other subbands makes it difficult to evaluate the intercept accurately.

Further low-temperature quantum transport was also measured using a dilution refrigerator. In Supplementary Fig. 12, $R_{\mathrm{xx}}$ and $R_{\mathrm{yx}}$ taken at 40 mK for the 12 nm film are compared to 2 K ones, showing no major difference between them. So far, no more fine structures such as fractional quantum Hall states are confirmed in the present samples even at this ultra-low temperature up to 14 T.

**Construction of electronic structure.**

The band gap $E_{\mathrm{g}}$ is also estimated from the $g$ factor change in the following Roth equation, which is derived in the second order of the $k \cdot p$ perturbation theory [30].

$$g = 2 - \frac{2}{3} \frac{E_{\mathrm{p}} \Delta}{E_{\mathrm{g}}(E_{\mathrm{g}} + \Delta)} \tag{7}$$

Here $\Delta$ is the spin-orbit splitting energy (0.27 eV) [29] and $E_{\mathrm{p}}$ is the energy equivalent of the principal interband momentum matrix element ($-0.68$ eV). $E_{\mathrm{g}}$ is estimated at $\gtrsim 110$, $\gtrsim 110$, $\sim 55$, and $\sim 30$ meV for the 12, 14, 16 and 23 nm films, by using the above parameters and respective $g$ factors ($\lesssim 5$, $\lesssim 5$, $\sim 9$, and $\sim 15$). These electronic structures parameters are summarized in Table 1.

The total phase shift at both interfaces was estimated as a function of the binding energy $E_B$ by assigning band gap values $E_g$ and chemical potential differences $\Delta\phi$ of SrTiO$_3$, TiO$_2$, and Cd$_3$As$_2$ to the following empirical formula [44].

$$
\begin{aligned}
\delta(E) &= 2\arcsin\sqrt{\frac{E_{g,\mathrm{SrTiO_3}} - \Delta\phi_{\mathrm{SrTiO_3,Cd_3As_2}} - E_B}{E_{g,\mathrm{SrTiO_3}}}} \\
&+ 2\arcsin\sqrt{\frac{E_{g,\mathrm{TiO_2}} - \Delta\phi_{\mathrm{TiO_2,Cd_3As_2}} - E_B}{E_{g,\mathrm{TiO_2}}}} - 2\pi
\end{aligned}
\tag{8}
$$

**Data availability.**

The data supporting the plots within the paper and its Supplementary Information File are available from the corresponding author upon reasonable request.

Table 1: Parameters of electronic structures for Cd$_3$As$_2$ films of various thicknesses.

| $t$ (nm) | $m^*$ ($m_0$) | $E_F$ (meV) | $k_F$ (Å$^{-1}$) | $g$ | $E_g$ (meV) |
|---|---|---|---|---|---|
| 12 | 0.042 | 116 | 0.032 | $\lesssim 5$ | $\gtrsim 110$ |
| 14 | 0.038 | 129 | 0.032 | $\lesssim 5$ | $\gtrsim 110$ |
| 16 | 0.038 | 114 | 0.031, 0.023 | $\sim 9$ | $\sim 55$ |
| 23 | 0.035 | 143 | 0.032, 0.026 | $\sim 15$ | $\sim 30$ |
| 100 (bulk) | 0.049 | 134 | 0.037 | $\sim 15$ | 0 |

**Acknowledgements**

We acknowledge fruitful discussions with R. Arita, T. Koretsune, S. Nakosai, R. Yoshimi, M. Kawamura, K. Muraki, K. Ishizaka, M. Hirschberger, and T. Liang. We also thank M. Tanaka, S. Ohya, and A. Kikkawa for technical advice about the handling of arsenides. This work was supported by JST CREST Grant No. JPMJCR16F1, Japan and by Grant-in-Aids for Scientific Research (S) No. JP24226002, Scientific Research (C) No. JP15K05140, Young Scientists (A) No. JP15H05425, and Scientific Research on Innovative Areas "Topological Materials Science" No. JP16H00980 from MEXT, Japan.

**Author contributions**

M.U. and M.Kawasaki conceived the project. M.U., Y.N. and S.N. synthesized the bulk targets with M.Kriener and grew the thin films. M.U., Y.N., K.A., A.M. and M.T. performed the high-field measurements. M.U. analysed the data and wrote the manuscript with input from all authors. Y.K., Y.Taguchi, M.T., N.N., Y.Tokura and M.Kawasaki jointly discussed the results.

**Additional information**

Supplementary information is available in the online version of the paper. Reprints and permissions information is available online at www.nature.com/reprints. Correspondence and requests for materials should be addressed to M.U.

**Competing financial interests**

The authors declare no competing financial interests.

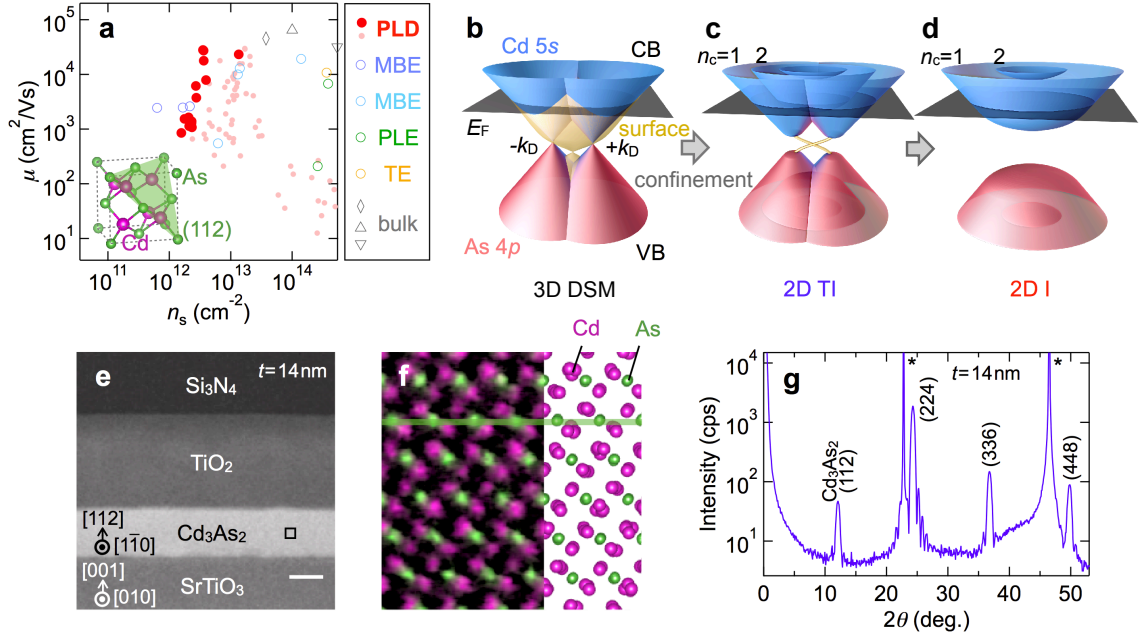

Figure 1: **High-crystallinity and high-mobility Cd₃As₂ thin films. a**, Experimental trend of electron mobility versus sheet carrier density. Among our films (●) prepared with pulsed laser deposition (PLD), high-quality ones obtained by high-temperature annealing are highlighted with bigger symbols. The mobility reaches a maximum of $\mu = 3 \times 10^4$ cm²/Vs even at a thickness of $t = 30$ nm, rivaling mobility values for bulk thinned plates ($\diamondsuit$ [22]) and nanostructures ($\triangle$ [23] and $\triangledown$ [24]), while it intrinsically decreases with reducing to two dimensions. Other films (○) grown by molecular beam epitaxy (MBE) [18, 19], thermal evaporation (TE) [15], or pulsed laser evaporation (PLE) [16] techniques are also plotted for comparison. Inset shows the primary cubic structure of Cd₃As₂. **b-d**, Schematic illustration of the electronic structure evolution from the 3D DSM state. With decreasing the film thickness, subbands are formed due to the quantum confinement, giving rise to two-dimensional topological insulating (2D TI) and trivial insulating (2D I) states depending on the number of inverted subbands [6]. **e**, Cross-sectional image of a 14 nm Cd₃As₂ film sandwiched between Si₃N₄/TiO₂ cap and SrTiO₃ substrate. The length of the scale bar is 10 nm. **f**, Atomically resolved element map of the boxed region in **e**, shown with the cross-sectional view of the crystal structure. **g**, In the x-ray diffraction $\theta-2\theta$ scan, Bragg peaks of the (112)-oriented Cd₃As₂ film are observed with clear Laue oscillations. The SrTiO₃ substrate peaks are marked with an asterisk.

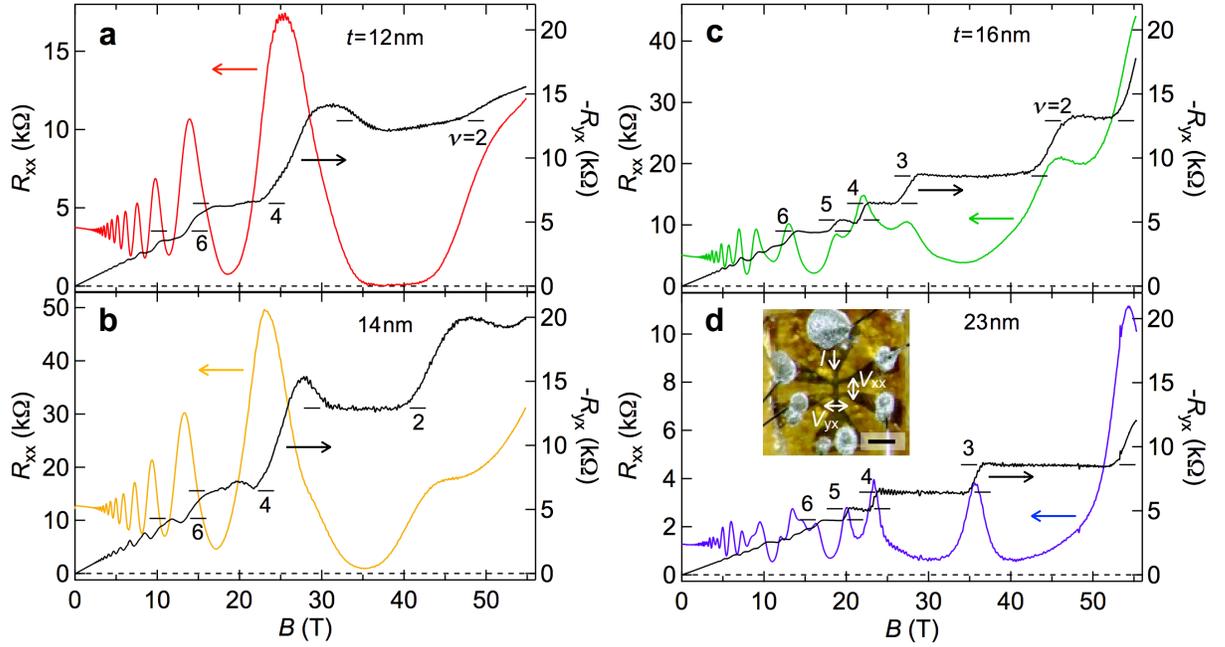

Figure 2: **Quantum Hall effect observed in Cd$_3$As$_2$ films. a** and **b**, High-field magnetotransport for thin films ($t = 12$ and $14$ nm) measured at $T = 1.4$ K. The numbers of the horizontal bars represent the filling factor $\nu$. The degeneracy factor $s$ is determined to be $s = 2$ from the increment of the plateau values. **c** and **d**, Same scan for slightly thicker films ($t = 16$ and $23$ nm). By contrast, the degeneracy factor is altered to $s = 1$ in these thicker films at high fields. Inset depicts a measured Hall bar with a channel width of $60$ $\mu$m. The length of the scale bar is $300$ $\mu$m.

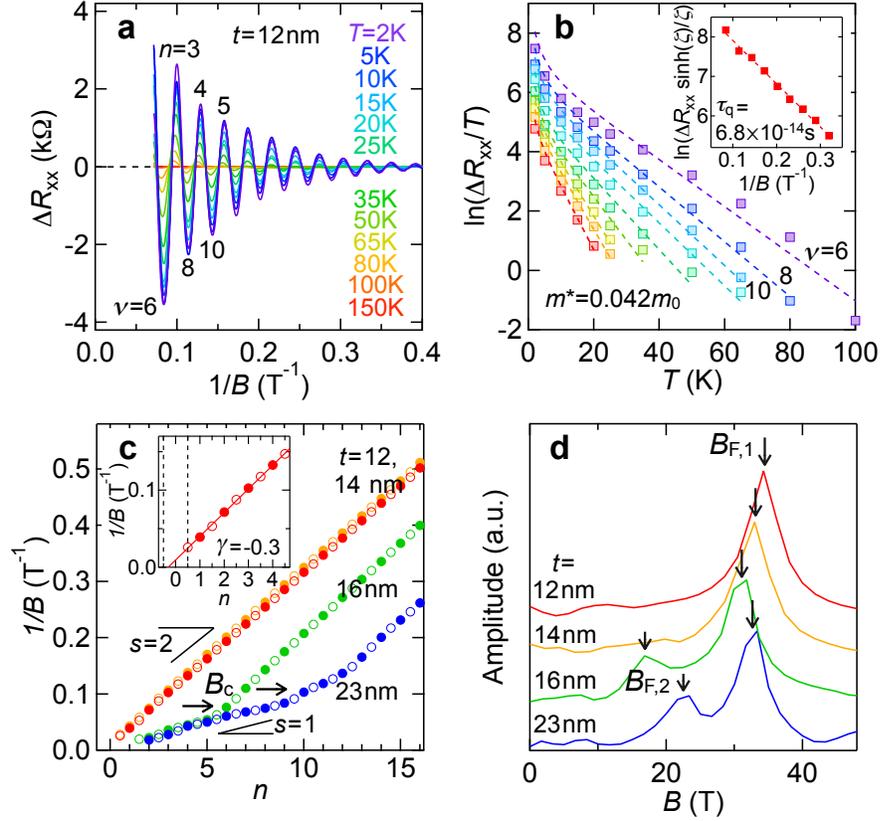

Figure 3: **Thickness-dependent quantum transport characteristics. a**, Temperature dependence of the SdH oscillations for the $12$ nm $Cd_3As_2$ film, plotted against $1/B$ after subtracting a smooth background from $R_{xx}$. **b**, Dingle analysis of the oscillation amplitude to obtain effective mass and quantum scattering time (inset). **c**, Landau-level fan diagram plotted for the series of films with different thicknesses, and its magnification to evaluate the intercept for the $12$ nm film (inset). The integer (half-integer) indices at $R_{xx}$ peak (valley) are denoted by a closed (open) circles. In the thicker 16 nm and 23 nm films, spin splitting of the oscillations occurs above the critical field $B_c$. **d**, Fourier transformation of the SdH oscillations below 12 T to extract the oscillation frequencies $B_{F,1}$ and $B_{F,2}$.

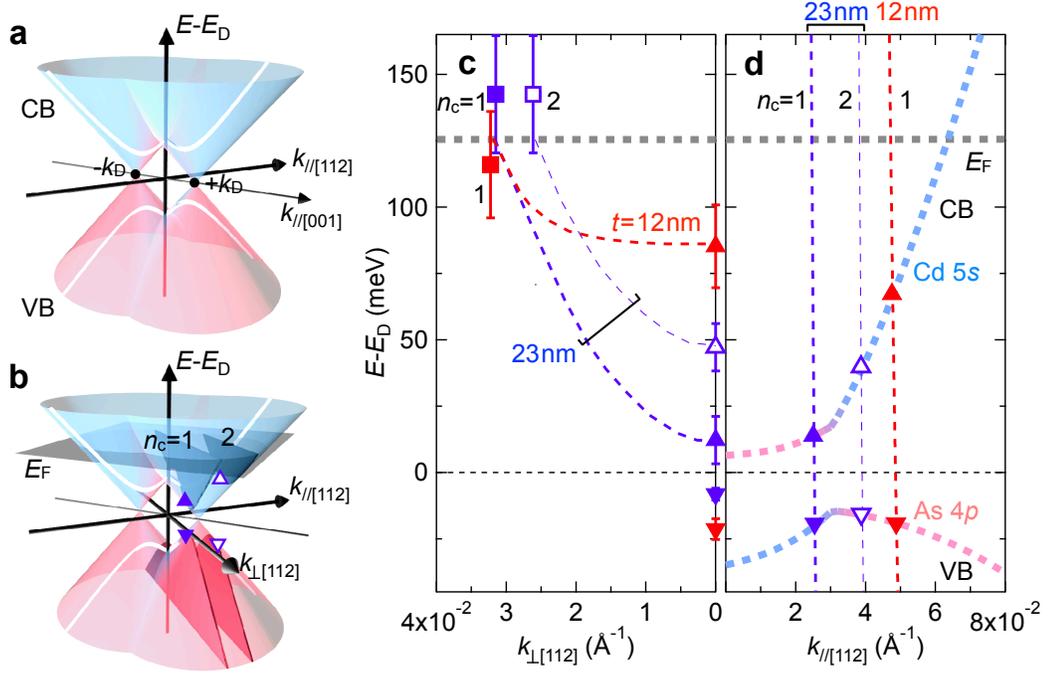

Figure 4: **Electronic structures identified from quantum transport. a**, Cross sectional dispersions along the out-of-plane field direction ($k_{\parallel[112]}$). **b**, Subband formation ($n_c = 1$ and 2) within the in-plane direction ($k_{\perp[112]}$), for $t = 23$ nm. **c** and **d**, Constructed electronic structures measured from the Dirac point energy $E_D$, are shown for the two typical thicknesses ($t = 12$ and 23 nm). (see Supplementary Fig. 13 for all the thicknesses). In **c**, the obtained Fermi energy $E_F$ and in-plane Fermi momentum $k_F$ are plotted for the split subbands (square). The bottom of the conduction band (CB) and the top of the valence band (VB) interpolated from previous calculations [17] are indicated by an upward and a downward triangle, respectively. The error bars for $E_F$ were estimated from Eq.1 by assigning effective mass values obtained using the Lifshitz-Kosevich formula for each filling state $\nu$ (see Fig. 3b). The structural plot in **d** represents the Bohr-Sommerfeld quantisation condition along the film normal direction (Eqs. 2 and 3), where the crossing point (triangle) of the original calculated band dispersions (dotted lines, left-hand side of Eq. 3) [6] and the quantization condition curves (vertical dashed lines, right-hand side of Eq. 3) determines the energy and momentum of the bottom of the split subbands. The orbital character of each subband is also identified, indicating the topological phase transition as represented by the color change of the subband ($n_c = 1$) from blue (Cd $5s$) to pink (As $4p$). Subband data of $n_c = 1$ (2) are distinguished by closed (open) symbols and thick (thin) dashed lines.

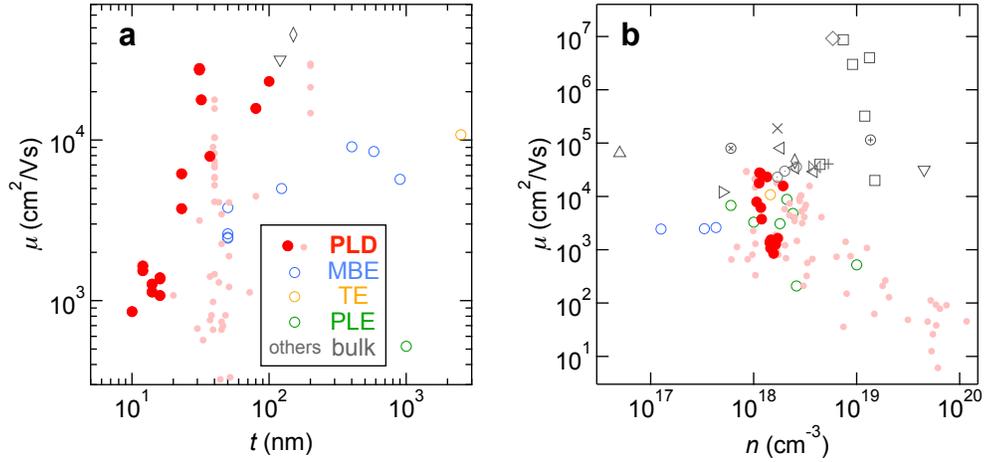

Supplementary Figure 1: **Trends of mobility and carrier density in Cd₃As₂ samples.** Electron mobility versus (**a**) sample thickness and (**b**) volume carrier density summarized for Cd₃As₂ film and bulk samples fabricated to date. Our films prepared with pulsed laser deposition (PLD) are indicated by closed circles, and in particular, high quality ones obtained by high temperature annealing are highlighted with bigger circles. Other films previously grown by molecular beam epitaxy (MBE) [1, 2], thermal evaporation (TE) [3], or pulsed laser evaporation (PLE) [4, 5] techniques are represented by open circles. Bulk samples grown such as by Cd flux or chemical vapor transport are denoted by other gray symbols (◇ [6], △ [7], ▽ [8], □ [9], ◇ [10], ◁ [11], ▷ [12], ⋈ [13], × [14], + [15], ⊙ [16], ⊗ [17], and ⊕ [18]).

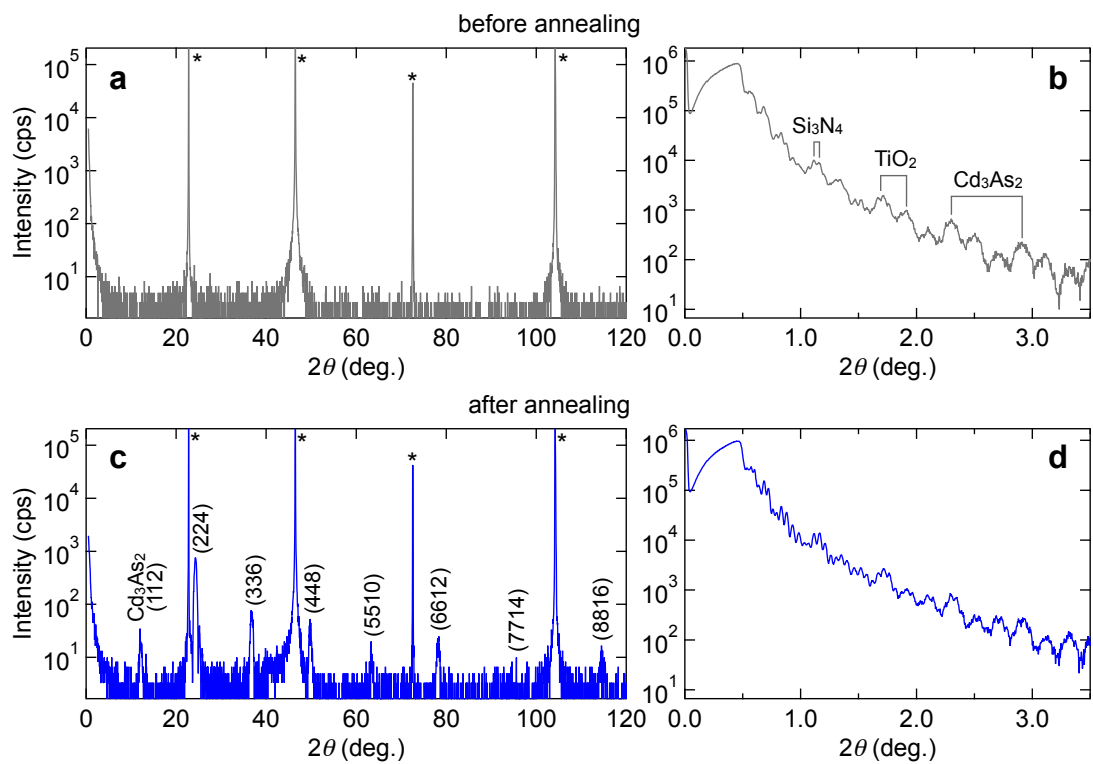

Supplementary Figure 2: **Annealing effect probed by x-ray diffraction.** (**a**) Typical x-ray diffraction $\theta$–$2\theta$ scan and (**b**) small angle reflectivity curve of a $Cd_3As_2$ film capped with $TiO_2/Si_3N_4$ layers before annealing. The $SrTiO_3$ substrate peaks are marked with an asterisk. (**c**) and (**d**) Results of the same film after high-temperature annealing.

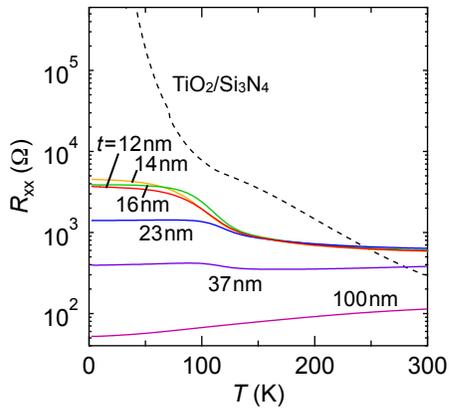

Supplementary Figure 3: **Temperature dependence of the sample resistance.** Temperature dependence of $R_{xx}$ is plotted for the capped $Cd_3As_2$ films with various thicknesses, also compared to a sample consisting only of the $TiO_2$ / $Si_3N_4$ capping layers.

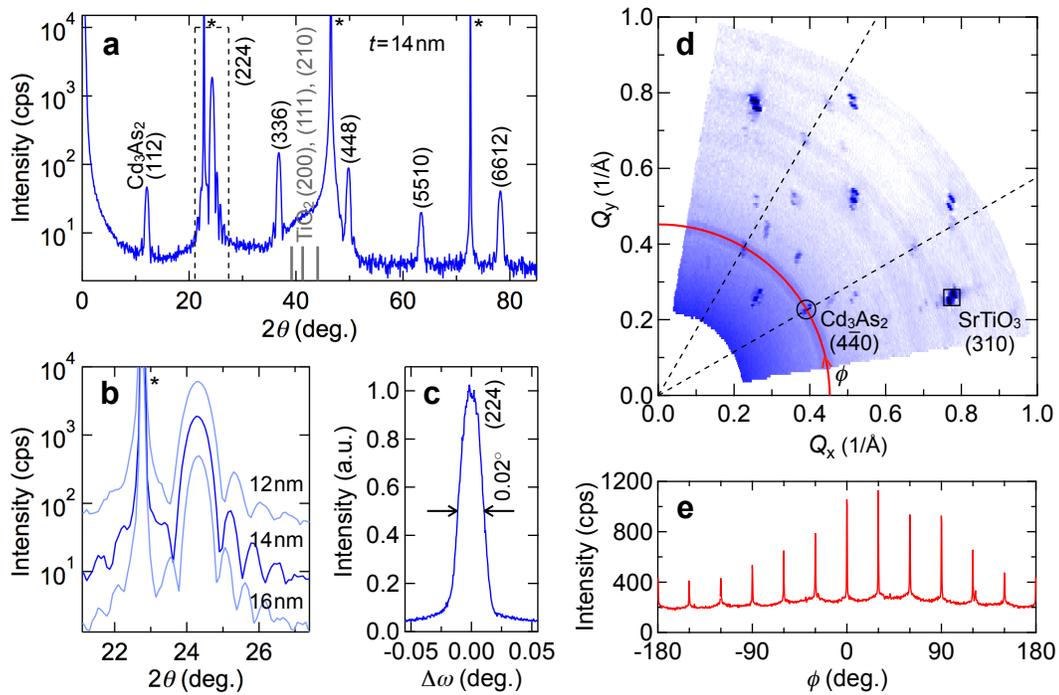

Supplementary Figure 4: **X-ray diffraction characterization of the Cd$_3$As$_2$ films.** (**a**) Lower speed $\theta$–$2\theta$ scan of the 14 nm Cd$_3$As$_2$ film. The substrate peaks are marked with an asterisk. A background hump around 40 degrees is ascribable to polycrystalline TiO$_2$ peaks. (**b**) Magnification of the (224) peak and its Laue oscillations, compared to those for 12 and 16 nm films. (**c**) Rocking curve of the (224) film peak with a full width at half maximum (FWHM) of 0.02 degrees. (**d**) In-plane reciprocal space mapping and (**e**) $\phi$ scan of the ($4\bar{4}0$) peak, showing locked in-plane orientation of the film.

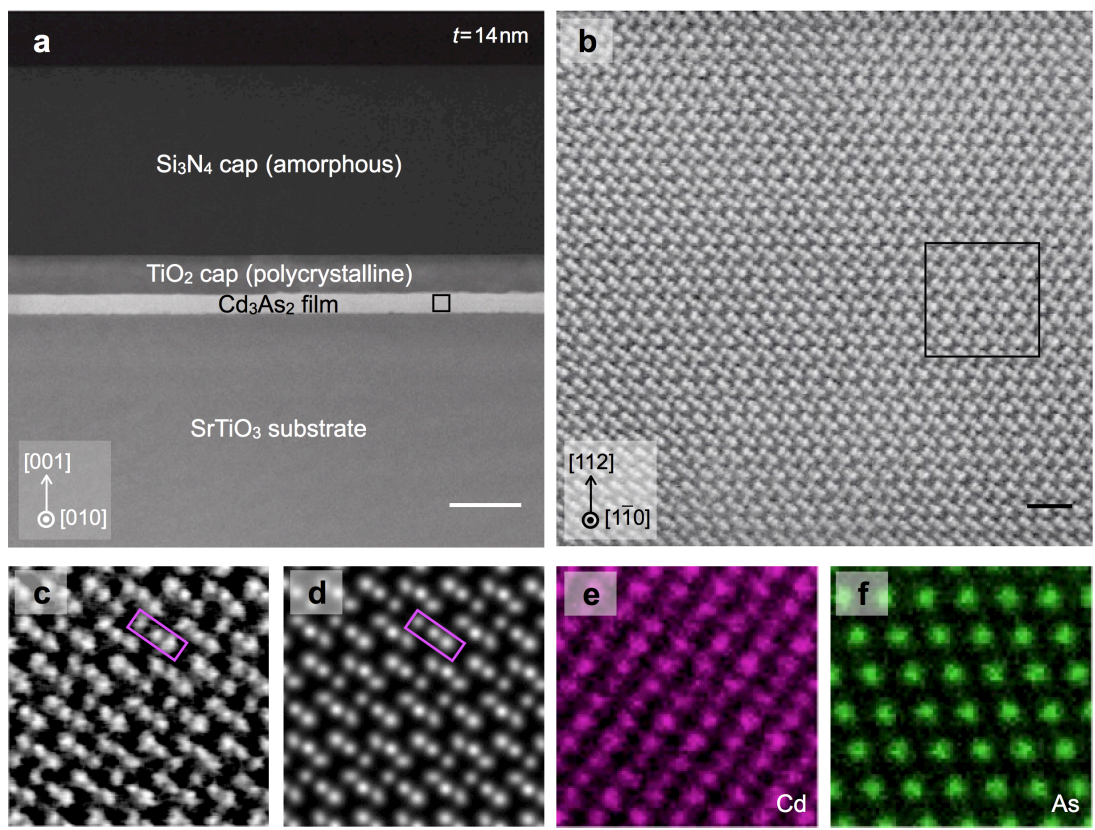

Supplementary Figure 5: **Transmission electron microscopy characterization of the Cd$_3$As$_2$ film.** (**a**) Overall picture of the heterostructure and (**b**) magnified view of the Cd$_3$As$_2$ film of the boxed area in (**a**), taken with cross-section high-angle annular dark-field scanning transmission electron microscopy (HAADF-STEM). The lengths of the scale bars are 50 nm and 1 nm, respectively. (**c**) Higher-resolution magnified image of the Cd$_3$As$_2$ film of the boxed area in (**b**), (**d**) simulated image for the low-temperature structure of Cd$_3$As$_2$ ($I4_1/acd$), and corresponding element maps taken with energy dispersive x-ray spectrometry (EDX) for (**e**) Cd $L$ and (**f**) As $K$ edges. Shift of Cd atoms peculiar to the low-temperature structure is clearly confirmed as indicated in the box.

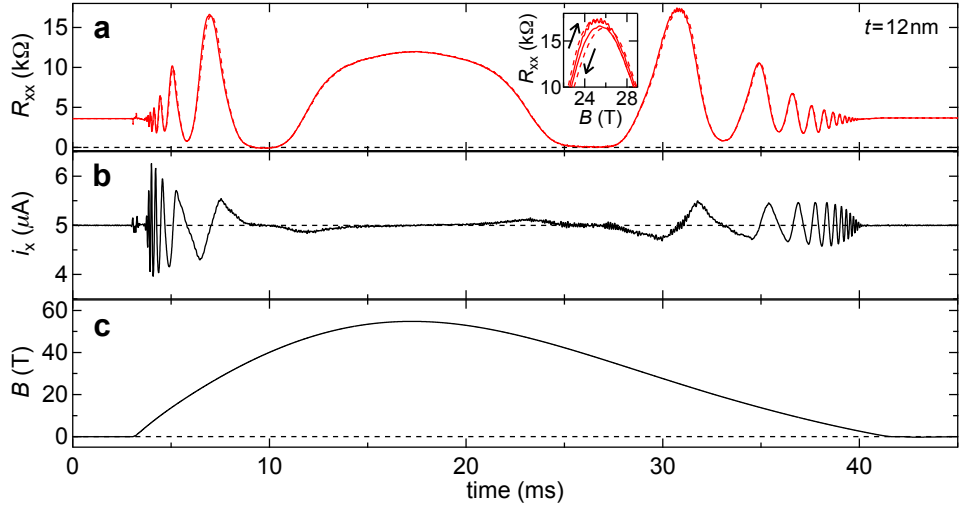

Supplementary Figure 6: **Correction of data taken in the pulsed magnetic fields.** (**a**) Raw (dashed line) and corrected (solid line) $R_{xx}$ curve for the 12 nm film, taken in a single scan of pulsed field shown in (**c**). The inset shows magnified data resolving hysteresis. The correction was made using a classic model [19] as expressed in Eq. S1, in which a small capacitive component of 5 nF is taken into consideration to calculate (**b**) a time-variable actual current $i_x$ though the sample.

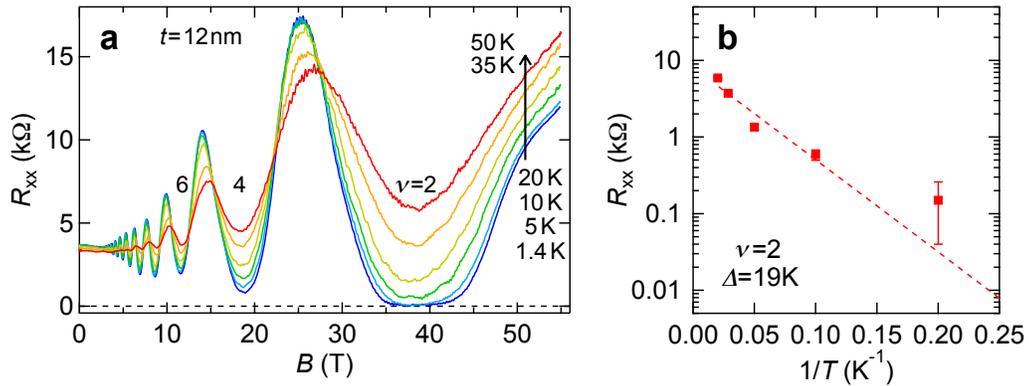

Supplementary Figure 7: **Temperature dependence of quantum Hall effect.** (**a**) Temperature dependence of $R_{xx}$ for the 12 nm $Cd_3As_2$ film in high magnetic fields. (**b**) Arrhenius plot of the $R_{xx}$ minima for the $\nu = 2$ quantum Hall state. The error bars were estimated from maximum and minimum values in the upward and downward sweeps.

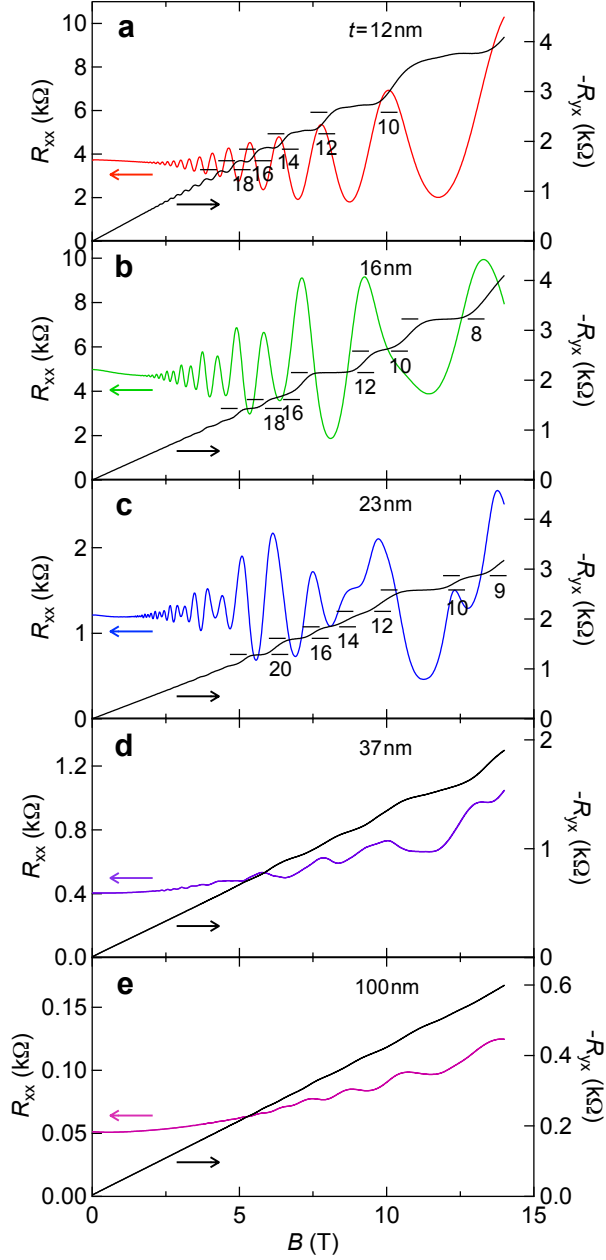

Supplementary Figure 8: **Low-field magnetotransport dependent on the confinement thickness.** $R_{xx}$ and $R_{yx}$, taken using a conventional superconducting magnet at 2 K, are plotted for the (**a**) 12 nm, (**b**) 16 nm, (**c**) 23 nm, (**d**) 37 nm, and (**e**) 100 nm films. Shubnikov-de Haas (SdH) oscillations are resolved for a field as low as a few teslas, and corresponding clear Hall plateaus are also confirmed for the films below 23 nm. The numbers with the horizontal bars represent -1/$R_{yx}$ values in the unit of $e^2/h$.

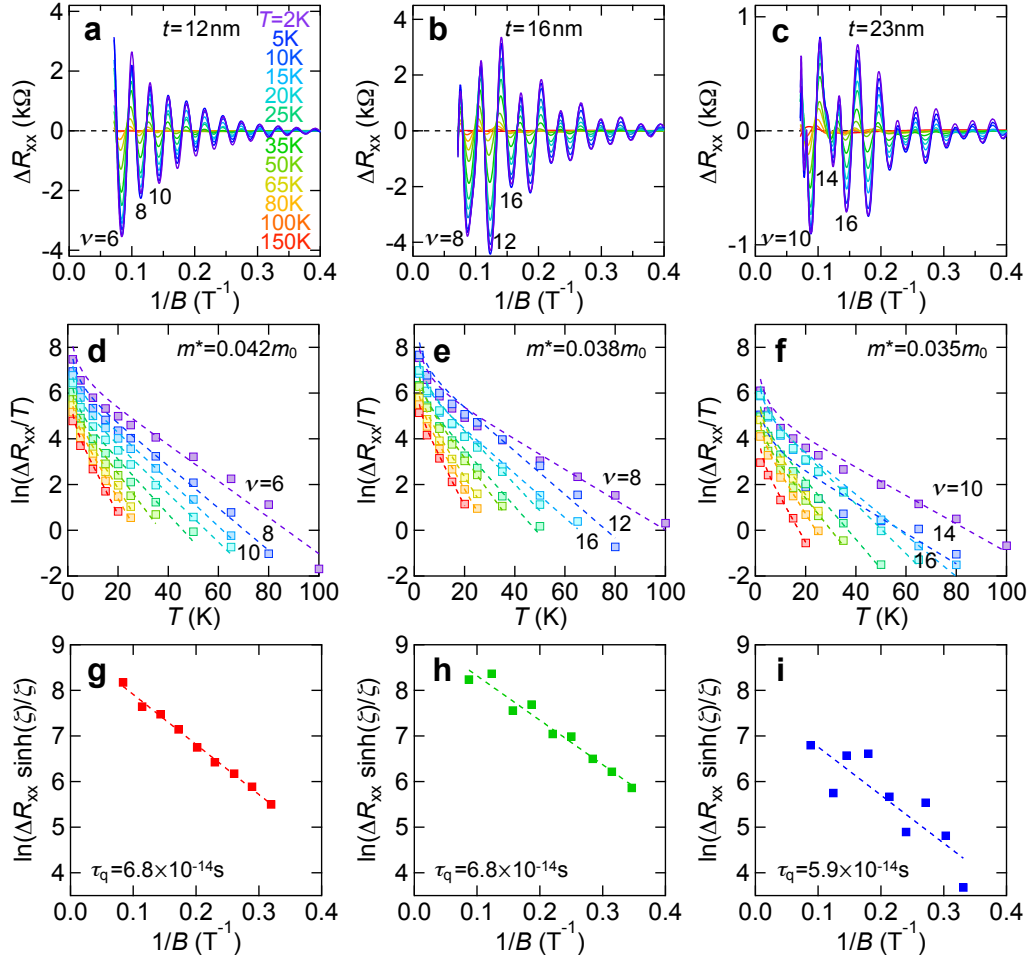

Supplementary Figure 9: **Temperature dependence of the SdH oscillations and their analysis.** (**a**)-(**c**) Temperature dependence of the SdH oscillations, plotted as a function of $1/B$ after subtracting a smooth background. (**d**)-(**f**) Lifshitz-Kosevich analysis of the oscillation amplitude for each filling state. The broken curves are fits to the formula, yielding an effective mass of $0.035 \sim 0.042 m_0$. (**g**)-(**i**) Dingle plots of the oscillation amplitude. A quantum scattering time of $\tau_q = 6 \sim 7 \times 10^{-14}$ s is extracted from the slope of the linear fit, corresponding to Dingle temperature of $T_D = 18 \sim 21$ K.

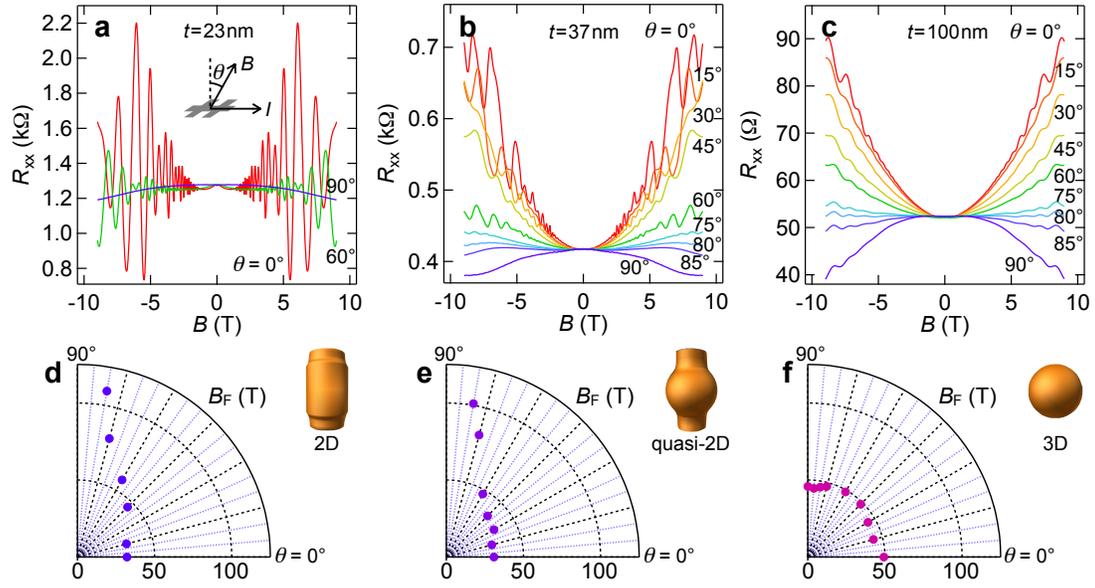

Supplementary Figure 10: **Field angle dependence of the SdH oscillations reflecting the dimensional change.** Evolution of the SdH oscillations measured with changing the field direction from $\theta = 0°$ ($B \perp I$) to $90°$ ($B \parallel I$) for the (**a**) 23 nm, (**b**) 37 nm, and (**c**) 100 nm films at 2 K. (**d**)-(**f**) Polar plots showing the angle dependence of the oscillation frequency for the three films. As contrasted to the two-dimensional (2D) behavior in the 23 nm film, quasi-2D behavior appears at the 37 nm film and it changes into three-dimensional (3D) at the 100 nm film.

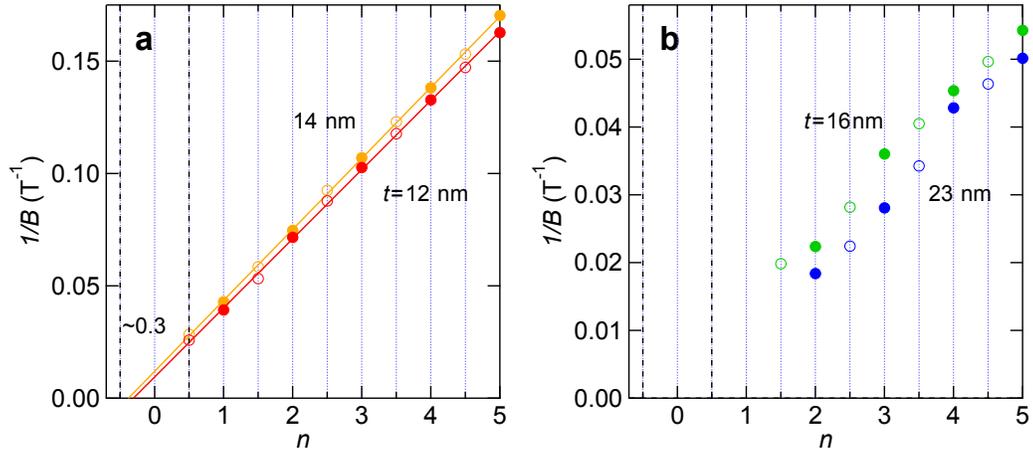

Supplementary Figure 11: **Landau-level fan diagrams magnified around the origins.** Landau-level fan diagrams plotted for (**a**) the thinner ($t = 12$, 14 nm) and (**b**) thicker ($t = 16$, 23 nm) films. The integer (half-integer) indices at $R_{xx}$ peak (valley) are denoted by a closed (open) circles. For the 16 nm and 23 nm films, spin splitting of the oscillations already occurs in this field range.

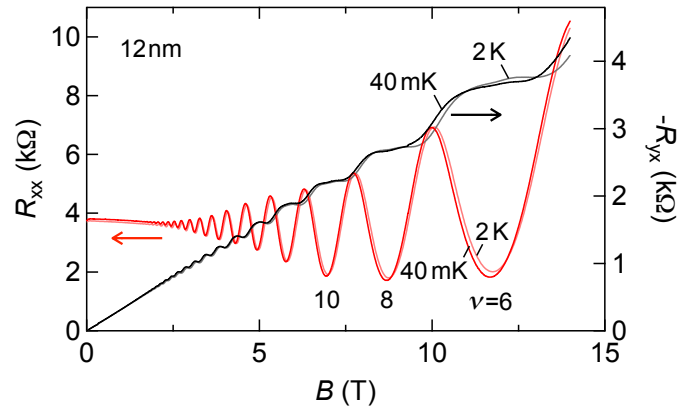

Supplementary Figure 12: **Ultra-low temperature magnetotransport measured at 40 mK.** $R_{xx}$ and $R_{yx}$, taken using a dilution refrigerator with the base temperature, are plotted for the 12 nm $Cd_3As_2$ film. It does not show any significant difference from that at 2 K.

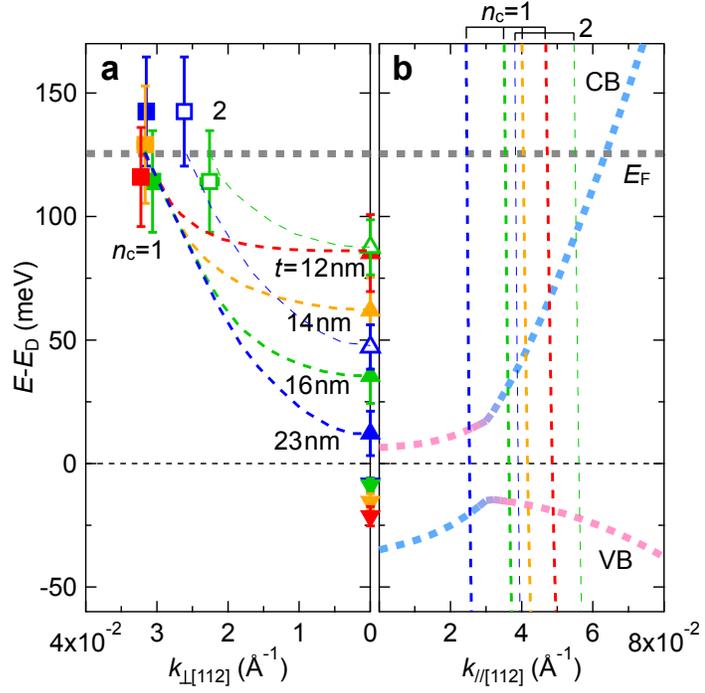

Supplementary Figure 13: **Electronic structures under the quantum confinement.** Constructed electronic structures are plotted for all the thicknesses ($t = 12, 14, 16,$ and 23 nm), while in Fig. 4 in the main text they are shown only for the two typical thicknesses ($t = 12$ and 23 nm) for clarity. In (**a**), the Femi energy and in-plane momentum ($k_{\perp[112]}$) obtained from the quantum transport analysis are denoted by a square. The bottom of the conduction band (CB) and the top of the valence band (VB) are indicated by an upward and a downward triangle. Subband data ($n_c = 1$ and 2) are distinguished by closed and open symbols, respectively. (**b**) Quantum confinement condition along the film normal direction ($k_{\parallel[112]}$).

# Supplementary References